\documentclass[12pt,preprint]{aastex}

\begin{document}

\title{Extragalactic Radio Sources and the WMAP Cold spot}
\author{Lawrence Rudnick \footnote{larry@astro.umn.edu}, Shea Brown\footnote{brown@astro.umn.edu}, Liliya R. Williams\footnote {llrw@astro.umn.edu}}
\affil{Department of Astronomy, University of Minnesota\\ 116 Church St. SE,  Minneapolis, MN  55455}

\begin{abstract}
We detect a  dip of 20-45$\%$ in the surface brightness and number counts of NVSS sources smoothed to a few degrees  at the location of the WMAP cold spot.   The dip has structure on scales of $\sim$ 1$^\circ$ to 10$^\circ$.  Together with independent all-sky wavelet analyses, our results suggest that  the dip in extragalactic brightness and number counts and the WMAP cold spot are physically related, i.e., that the coincidence is neither a statistical anomaly nor a WMAP foreground correction problem. If the cold spot does originate from structures at modest redshifts, as we suggest, then  there is no remaining need for non-Gaussian processes at the last scattering surface of the CMB to explain the cold spot.  The late integrated Sachs-Wolfe effect, already seen statistically for NVSS source counts, can now  be seen to operate on a single region.  To create the magnitude and angular size of the  WMAP cold spot requires a $\sim$~140 Mpc radius completely empty void at z$\leq$1 along this line of sight.  This is far outside the current expectations of the concordance cosmology, and adds to the anomalies seen in the CMB.

\end{abstract}

\section{Introduction}

The detection of an extreme ``cold spot'' \citep{v04}  in the foreground-corrected WMAP images was an exciting but unexpected finding. At 4$^\circ$ resolution, Cruz et al. (2005) determine an amplitude of -73~$\mu$K, which reduces to -20~$\mu$K at $\sim$10$^\circ$ scales \citep{cruz07}.  The non-gaussianity of this extreme region has been scrutinized, (Cayon, Jun \& Treaster 2005; Cruz et al. 2005, 2006, 2007)   concluding that it cannot be explained by either foreground correction problems  or  the normal Gaussian fluctuations of the CMB.  Thus, the cold spot seems to require  a distinct origin -- either primordial or local.  Across the whole sky, local mass tracers such as the optical Sloan Digital Sky Survey \citep[SDSS,][]{york}  and the radio NRAO VLA Sky Survey 
\citep[NVSS,][]{nvss} are seen to correlate with the WMAP images of the CMB \citep{piet06,cabre06}, probably through the late integrated Sachs-Wolfe effect \citep[ISW,][]{crit1}.

\cite{MCE} extended the study of radio source/CMB correlations by performing a steerable wavelet analysis of NVSS source counts and WMAP images.   They isolated 18 regions that, as a group, contributed a significant fraction of the total NVSS-ISW signal.  Three of those 18 regions were additionally robust to the choice of wavelet form.   The centroid of one of those three robust correlated regions, (\#16),  is inside the 10$^\circ$ cold spot derived from WMAP data alone \citep{cruz07}, although \cite{MCE} did not point out this association.  

The investigations reported here were conducted independently and originally without knowledge of the \cite{MCE} analysis. However, our work is {\it a posteriori} in nature, because we were specifically looking for the properties in the direction of the cold spot.  These results thus support and quantify the NVSS properties in the specific direction of the cold spot, but should be considered along with the \cite{MCE} analysis for the purposes of an unbiased proof of a WMAP association.

\section{Analysis and Characterization of the NVSS ``dip''}

We examined both the {\it number counts} of NVSS sources in the direction of the WMAP cold spot and their {\it smoothed brightness distribution}.  The NVSS 21~cm survey  covers the sky above a declination of -40$^\circ$ at a resolution of 45''. It has an rms noise of 0.45 mJy/beam and is accompanied by a catalog of sources stronger than 2.5 mJy/beam.  Because of the short interferometric observations that went into its construction, the survey is insensitive to diffuse sources greater than $\approx$ 15' in extent. Convolution of the NVSS images to larger beam sizes, as done here, shows the integrated surface brightness of small extragalactic sources;  this is very different than what would be observed by a single dish of equivalent resolution.  In the latter case, the diffuse structure of the Milky Way Galaxy dominates (e.g, \cite{haslam81}), although it is largely invisible to the interferometer.

To explore the extragalactic radio source population in the direction of the WMAP cold spot, we first show in Figure 1 the 50$^\circ\times$50$^\circ$ region around the cold spot convolved to a resolution of  3.4$^\circ$.  Here, the region of the cold spot is seen to be the faintest region on the image (minimum at l$_{II}$,b$_{II}$~=~ 207.8$^\circ$,~-56.3$^\circ$).  At minimum, its brightness is 14~mK below the mean, with an extent of $\approx$5$^\circ$.
The WMAP cold spot thus picks out a special region in the NVSS -- at least within this 2500$^\square$ region.

We examined the smoothed brightness distribution across the whole NVSS survey using another averaging technique that reduces the confusion from the brightest sources. We first pre-convolved the images to 800'', which fills all the gaps between neighboring sources, and then calculated the {\it median} brightness in sliding boxes 3.4$^\circ$ on a side.   The resulting image is shown in Figure \ref{allsky} which is in galactic coordinates centered at  l$_{II}$=180$^\circ$.   The dark regions near the galactic plane are regions of the NVSS survey that were perturbed by the presence of very strong sources. Note that the galactic plane itself, which dominates  single dish maps,  is only detectable here between -20$^\circ < l_{II} < $ 90$^\circ$, where there is a local increase in the number of small sources detectable by the interferometer.

To evaluate the NVSS brightness properties of the cold spot, we compared it with the  distribution of median brightnesses in two strips from this all sky map.  The first strip was in the north, taking everything above a nominal galactic latitude of 30$^\circ$ (More precisely, we used the horizontal line in the Aitoff projection tangent to the 30$^\circ$ line at  l$_{II}$=180$^\circ$.)  The second strip was in the south, taking everything below a nominal galactic latitude of -30$^\circ$, but only from  10$^\circ  < l_{II}< $180$^\circ$, to avoid regions near the survey limit of $\delta$=-40$^\circ$. The minimum brightness in the cold spot region ($\approx$~20~mK) is equal to the lowest values seen in the 16,800 square degree area of the two strips, and is $\approx$~30\% below the mean (Figure \ref{coldcum}). 

Formally, the probability of finding this weakest NVSS spot within the $\approx$10$^\circ$ (diameter) region of the WMAP cold spot is 0.6\%.  This {\it a posteriori} analysis thus is in agreement with the statistical conclusions of \cite{MCE} that the NVSS properties in this region are linked to those of WMAP.  We also note that the magnitude of the NVSS dip is at the extreme, but not an outlier of the overall brightness distribution. We thus expect that less extreme NVSS dips would also individually correlate with WMAP cold regions, although it may be more difficult to separate those from the primordial fluctuations.

The  NVSS brightness dip can be seen at a number of resolutions, and there is probably more than one scale size present.  At resolutions of 1$^\circ$, 3.4$^\circ$, and 10$^\circ$, we find that the dip is $\approx$ 60\%, 30\% and 10\% of the respective mean brightness.   At 10$^\circ$ resolution, the NVSS deficit overlaps with another faint region about 10$^\circ$ to the west, while the average dip in  brightness then  decreases from -14~mK (at 3.4$^\circ$)  to -4~mK.

The dip in NVSS brightness in the WMAP cold spot region  is not due to some peculiarity of the NVSS survey itself.  In Figure 4, we compare the 1$^\circ$ convolved NVSS image with the similar resolution, single dish 408 MHz image of \cite{haslam81}.   This 408 MHz all sky map  is dominated in most places by galactic emission, and was usedby \cite{1year03} as a template for estimating the synchrotron contribution in CMB observations.  On scales of  1$^\circ$, the {\it fluctuations} in brightness are a combination of galactic (diffuse) and extragalactic (smeared small source) contributions.  In the region of the cold spot, we can see the extragalactic contribution at 408 MHz by comparison with the smoothed NVSS 1.4 GHz image.  Note that although there is flux everywhere in the NVSS image, this is the ``confusion'' from the smoothed contribution of multiple small extragalactic sources in each beam, whereas the 408 MHz map has strong diffuse galactic emission as well.  Strong brightness dips are seen in both images in the region of the WMAP cold spot - with the brightness dropping by as much as 62\% in the smoothed NVSS;  in the 408 MHz map, this is diluted by galactic emission.

To look more quantitatively at the source density in the cold spot region, we measured the density of NVSS sources (independent of their fluxes) as a function of distance from the cold spot. Figure 5 shows the counts in equal area annuli around the WMAP cold spot down to two different flux limits.    
With a limit of 5 mJy, there is a 45$\pm$8\% decrease in counts in the 1$^\circ$ radius circle around the WMAP cold spot centroid. At the survey flux limit of 2.5 mJy, the decrease is 23$\pm$3\%.  This reduction in number counts is what is measured by \cite{MCE} in their statistical all-sky analysis.

\section{Foreground Corrections}

Several studies have claimed that the properties of the cold spot are most likely an effect of incorrect foreground subtraction 
\citep{cn04,coles05,lz05,toj06}.  This possibility has been investigated in detail for both the first year
 \citep{v04,c05,c06} and third \citep{cruz07} 
year WMAP data. The arguments {\it against} foreground subtraction
errors can be summarized in three main points -- 1) The region of the
spot shows no spectral dependence in the WMAP data.  This is
consistent with the CMB and inconsistent with the known spectral
behavior of galactic emission (as well as the SZ effect).  The flat
(CMB-like) spectrum is found both in temperature and kurtosis, as well
as in real and wavelet space. 2) Foreground emission is found to be
low in the region of the spot, making it unlikely that an
over-subtraction could produce an apparent non-Gaussianity.  3)
Similar results are found when using totally independent methods to
model and subtract out the foreground emission \citep{c06}, namely the
combined and foreground cleaned Q-V-W map \citep{1year03} and the
weighted internal linear combination analysis \citep{teg03}.

Now that we know that there is a reduction in the extragalactic radio source contribution in the direction of the cold spot, we can re-examine this issue.  We  ask whether a 20-30\% decrement in
the local brightness of the extragalactic synchrotron emission  could
translate into a foreground subtraction problem that could generate
the WMAP cold spot. We are not re-examining the foreground question
{\it ab initio}, simply examining the plausibility that the deficit of
NVSS sources could complicate the foreground calculations in this
location. The characteristic brightness in the 3.4$^\circ$ convolved
NVSS image around the cold spot is $\sim$ 51 mK at
1.4 GHz; the brightness of the cold spot is $\sim$ 37 mK. This
difference of 14 mK in brightness (4~mK at 10$^\circ$ resolution)
represents the extragalactic population contribution only, as the NVSS
is not sensitive to the large scale galactic synchrotron emission.  By
contrast, the single dish 1.4 GHz brightness within a few degrees of the
cold spot is $\sim$ 3.4~K, as measured using the Bonn Stockert 25~m
telescope \citep{bonn}.  Therefore the total synchrotron contribution at 1.4 GHz is
$\sim$~0.7~K above the CMB, 50 times larger than the localized extragalactic deficit.
 
One way to create the cold spot would be if the universal spectral
index used for the normal galactic (plus small extragalactic)
subtraction was incorrect for the {\it extra} brightness temperature contribution
of the NVSS dip,~ $\delta$T (-14~mK at L band, 1.4 GHz). We make
 an order of magnitude estimate of this potential
error. Following the first year data analysis \citep{1year03}, and the
similar exercise performed by \citet{c06}, we consider fitting a
synchrotron template map at some reference frequency $\nu_{ref}$, and
extrapolating the model, (F($\nu$)$_{model}$), with a spectral index of $\beta= -3$ to the Q,
V, and W bands. This  spectral index is consistent with those of \cite{c06}
and the average spectral index observed in the WMAP images
\citep{1year03,3year06}. Under the null hypothesis that the spectral index of $\delta$T
is the same as that of the mean brightness T$_0$, we then
calculate in the region of the deficit,

\begin{equation}F(\nu)_{model} = \left[T_0(\nu_{ref})+\delta~T(\nu_{ref})\right]\left(\nu/\nu_{ref}\right)^{-3}.\end{equation}

However, if the actual spectrum of $\delta$T is ~ -$\alpha$ (from L
band through W band) instead of -3, then the true foreground
subtraction, F$(\nu)_{true}$, should have been

\begin{equation}
F(\nu)_{true}=T_0\left(\nu_{ref}\right)\left(\nu/\nu_{ref}\right)^{-3}+\delta T(\nu_{ref})\left(\nu/\nu_{ref}\right)^{-\alpha}.
\end{equation} 

The foreground subtraction would then be in error  as a function of frequency as follows, expressed in terms of the L band temperatures :

\begin{equation}
\delta F\left(\nu\right) \equiv F(\nu)_{true}-F(\nu)_{model} = \delta T(\nu_L) (\nu_L/\nu)^{\alpha}\left[1-(\nu_{ref}/\nu)^{3-\alpha}\right].
\end{equation}

 Three different reference frequencies have been used for synchrotron extrapolation  -- the \citet{haslam81} 408
MHz map \citep{1year03}, the \cite{rhodes98} 2326 MHz Rhodes/HartRAO survey \citep{c06}, and the internal K
and Ka band WMAP images \citep{3year06}. 
Since the foreground subtraction errors would be worst extrapolating from the lowest frequency template, we start at 408~MHz and look at the problems caused by a spectral index for $\delta$T that is different than the assumed -3.  

We obtain a rough measure of the spectral index of the dip by comparing  the 1$^\circ$ resolution maps (Figure 4) at 408~MHz and 1400~MHz. At 
l$_{II}$,b$_{II}$~=~207$^\circ$,-55$^\circ$ , we find $\delta$T~=~2.6$\pm$0.75~K (30$\pm$12~mK) at 408 (1400) MHz, yielding a spectral index of -3.6$\pm$0.5. Using Equation (3),  this would actually lead to a WMAP ``hot spot'' if a spectral index of -3.0 had been assumed for the extrapolation.   Within the errors, the worst foreground extrapolation mistakes would then be hot spots that  range from +0.5 to +4.6~$\mu$K at Q band, 1$^\circ$ resolution, ($\approx$ +0.25 to +2.3~$\mu$K at 4$^\circ$ resolution). Our derived  spectral index for the dip is steeper than expected for extragalactic sources, so we also do the calculation assuming the flattest reasonable extragalactic spectrum of -2.5 .  This would lead to a spurious cold spot of -2.9~$\mu$K at 4$^\circ$ resolution .  In either case, this is far below the -73~$\mu$K observed at this resolution in WMAP, and  we thus conclude that the deficit of NVSS sources does not lead to a significant foreground subtraction error of either sign.

 \section{Discussion}

 The WMAP cold spot  could have three origins: a) at the last scattering
 surface ($z\sim 1000$), b) cosmologically local ($z<1$), or c) galactic.
 Because the spot corresponds to a significant deficit of flux (and source
 number counts) in the NVSS, we have argued here that the spot is
 cosmologically local and hence, a localized manifestation of the late ISW effect.

\cite{c05,cruz07}  derived a temperature deviation for the cold
spot of $\sim-20\mu$K
and a diameter  of $10^\circ$ using the WMAP 3 year data; on scales 
of $4^\circ$ the
average temperature is lower,  $\sim-73~\mu$K.  Using these two data
points we derive
an approximate relation between the temperature deviation and the
corresponding size of
the cold spot: $\theta(\Delta T/T)\approx 4.5\times 10^{-5}$, where
$\theta$ is the radius
of the cold spot in degrees.  We now  perform an order of magnitude
calculation to see if the
late ISW can  produce such a spot, assuming that the entire effect comes
from the ISW.

The contribution of the late ISW along a given line of sight is given by
$(\Delta~T/T)\vert_{ISW}~=-2\int\dot\Phi~d\eta$, where the dot represents
differentiation with respect to
the conformal time $\eta$, $d\eta=dt/a(t)$, and $a$ is the scale factor.
The integrand will be non-zero only at late times  (z$<$1) when the
cosmological
constant becomes dynamically dominant.

We start with the Newtonian potential given by
\begin{equation}
\Phi=GM/r\approx {{4\pi G}\over{3}}r^2\;\rho_b\;\delta.
\label{dphi}\end{equation}
The proper size $r$ and the background density $\rho_b$ scale as $a$ and
$a^{-3}$,
respectively. The growth of the fractional density excess, $\delta(a)$ in
the linear regime
is given by $D(a)=\delta(a)/\delta(a_0)$, and $D(a)$ is the linear growth
factor. For
redshifts below $\sim 1$ in $\Lambda$CDM, this factor can be
approximated as
$\delta(a)\approx a\delta(a_0)(3-a)/2$. Assuming that the region is
spherical, its
comoving radius is $r_c=0.5\Delta z(c/H)$, and $\Delta z$ is the line of
sight diameter
of the region. The change in the potential over $d\eta$ can be
approximated by
\begin{equation}
\Delta\Phi\sim{{4\pi G}\over{3}}r_{c}^2\;\rho_{c}\;\delta\;(\Delta z),
\end{equation}
where subscript $c$ refers to the average comoving size of the void and the
comoving background density. In  $\Lambda$CDM the Hubble parameter is
roughly
given by $H^2=H_0^2(1+2z)$, for redshifts below $\sim 1$. Incorporating
these
approximations we get the following relation between the size of the region, its
redshift and the temperature deviation from the late ISW:
\begin{equation} \label{dphi2}
\Delta\Phi\approx -{\Omega_m\over 2} \Bigl({{r_c}\over{c/H_0}}\Bigr)^3
                     (1+2z)^{1/2}(1+z)^{-2}\;\delta\approx {1\over 2}{{\Delta
T}\over T}
\end{equation}

We now ask under what conditions this expression is
 consistent with the observed relation between the size and temperature
of the cold spot
derived earlier, $\theta(\Delta T/T)\approx 4.5\times 10^{-5}$.  For
$\Omega_m\sim 0.3$
and $\delta=-1$ (i.e. a completely empty region) this leads to the
simplified relation,
$\theta(1+z)\approx 6$, where $\theta$ is in degrees, as before. Since the spot's association with the NVSS
places it at
$z\sim 0.5-1$, this leads to a self-consistent value of the radius of  $\sim 3 - 4^\circ$ for the observed spot. For
$c/H_0=4000$~Mpc,
the comoving radius of the void region is 120-160~Mpc.

How likely is such a large underdense region in a concordance cosmology?
Suppose
there is only one such large underdense region in the whole volume up to z=1.
The corresponding void frequency is then the ratio of the comoving volume
of the
void to the comoving volume of the Universe to z=1, which is roughly
$3\times 10^{-5}$.
Is this consistent with $\Lambda$CDM?
Void statistics have been done for a number of optical galaxy surveys,
as well as numerical structure formation simulations. Taking the most
optimistic
void statistics (filled dots in Fig.~9 of Hoyle \& Vogeley, 2004) which
can be
approximated by $\log P=-(r/{\rm Mpc})/15$, a 140~Mpc void would occur with a
probability
of $5\times 10 ^{-10}$, considerably more rare than our estimate for our
Universe
($3\times 10^{-5}$) based on the existence of the cold spot.
One must keep in mind, however, that observational and numerical void probability
studies  are limited to $r_c\sim 30$~Mpc;  it is not yet clear how these
should be extrapolated to $r_c > 100$~Mpc.

We note that \cite{in6a,in6b} had already suggested that
anomalous temperature anisotropies in the CMB, such as the cold spot,
may be explained by the ISW effect.  In contrast to our calculation
described above, their analysis considers the linear ISW plus the
second order effects due to an expanding compensated void, partially
filled with pressureless dust, embedded in a standard CDM (Inoue and Silk
2006a) or $\Lambda$CDM (Inoue and Silk 2006b) background. It is
reassuring that the size of the void indicated by their analysis---about
200 Mpc if located at $z\sim 1$---is roughly the same as what we get here
using linear ISW.
 
The need for an extraordinarily large void to explain the cold spot  would add  to the list of anomalies 
associated with the CMB. (See \cite{hol6a,hol6b} for a theory that
predicts such large voids based on a particular landscape model.)  These include
the systematically higher strength of the late ISW correlation measured for a variety of
mass tracers, compared to the WMAP predictions \citep[see Fig. 11 of][]{gian06}, and the
alignment and  planarity of the quadrupole and the octopole \citep{oliv,land}. We can,
however, conclude that models linking the cold spot with the larger scale anomalies, such as
the anisotropic Bianchi Type VII$_h$ model of Jaffe et al. (2005), are no longer necessary.
While we suggest that the  cold spot is a local effect, low order global anisotropic models \citep[e.g.,][]{pel06} may still be needed for the low$-\ell$ anomalies.

\section{Concluding Remarks}

We have detected a significant dip in the average surface brightness and number counts of radio sources from the NVSS survey at 1.4 GHz  in the direction of the WMAP cold spot. The deficit of extragalactic sources is also seen in a single dish image at 408 MHz.  Together with previous work, we rule out instrumental artifacts in WMAP due to foreground subtraction. A fuller examination of the statistical uncertainties associated with our combination of the \cite{MCE} wavelet results and  our own {\it a posteriori} analysis should be performed. With this caveat, we conclude that  the cold spot  arises from effects along the line of sight, and not at the last scattering surface itself. Any non-gaussianity of the WMAP cold spot therefore would then have  a local origin.

  A 140~Mpc radius, completely empty void at z$\le$1 is sufficient to create the magnitude and angular size of the cold spot through the late integrated Sachs-Wolfe effect.  Voids this large currently  seem improbable in the concordance cosmology, adding to the anomalies associated with the CMB.

  We suggest that a closer investigation of all  mass tracers would be useful to search for significant contributions from isolated regions.  Also, if our interpretation of the cold spot is correct,  it might be  possible to detect it indirectly using Planck, through the lack of lensing-induced
 polarization $B$ modes \citep{zald}.

\noindent{\bf ACKNOWLEDGMENTS}
\indent We thank Eric Greisen, NRAO, for improvements in the AIPS FLATN routine, which allows us to easily stitch together many fields in a flexible coordinate system. The 408 MHz maps were obtained through {\it SkyView}, operated under  the auspices of NASA's Goddard Space Flight Center. We appreciate discussions with M. Peloso, T. J. Jones and E. Greisen regarding this work, and useful criticisms from the anonymous referee.   LR acknowledges the  inspiration from his thesis adviser, the late David T. Wilkinson, who would have appreciated the notion of deriving information from a hole.   At the University of Minnesota, this work is supported in part,  through National Science Foundation grants AST~03-07604 and AST~06-07674 and STScI grant AR-10985.

\clearpage
\begin{figure}[t]
\begin{center}
\plotone{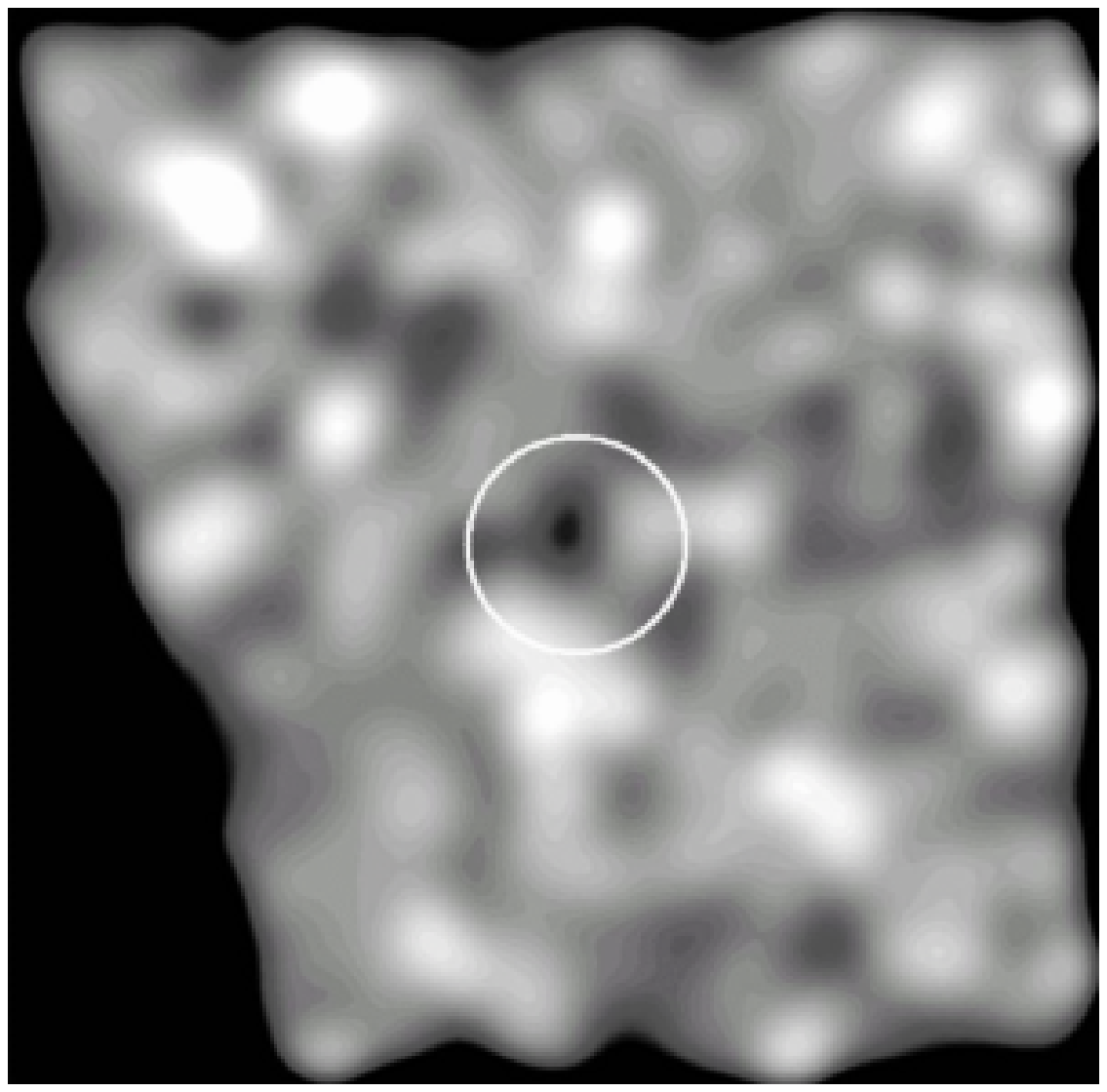}
\caption{ 50$^\circ$ field from smoothed NVSS survey at 3.4$^\circ$ resolution, centered at  l$_{II}$, b$_{II}$ = 209$^\circ$, -57$^\circ$. Values range from black: 9.3 mJy/beam to white: 21.5 mJy/beam.  A  10$^\circ$ diameter circle indicates the position and size of the WMAP cold spot.}
\label{cold3.4}\
\end{center}
\end{figure}

\begin{figure}[t]
\begin{center}
\plotone{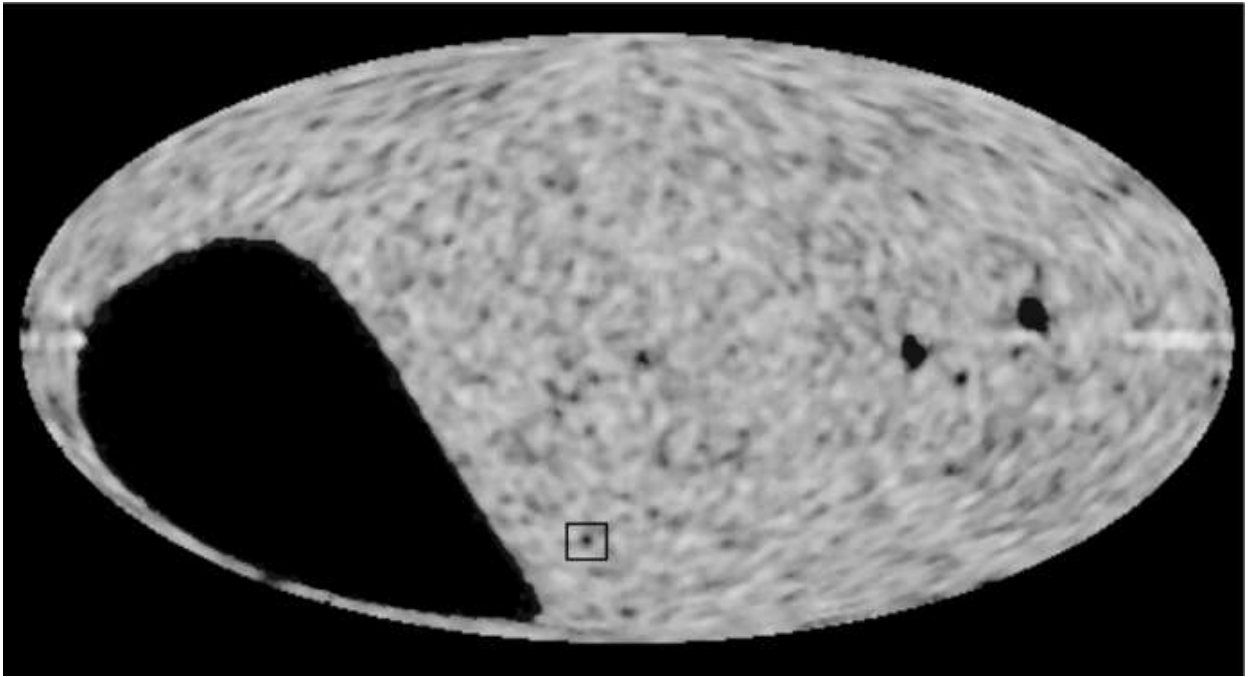}
\end{center}
\caption{Aitoff projection of NVSS survey, centered at l$_{II}$, b$_{II}$ = 180$^\circ$, 0$^\circ$, showing the median brightness in sliding boxes of 3.4$^\circ$. The WMAP cold spot is indicated by the black box. Closer to  the plane, large dark patches arise from sidelobes around strong NVSS sources. }
\label{allsky}
\end{figure}

\begin{figure}[t]
\begin{center}
\includegraphics[scale=0.21]{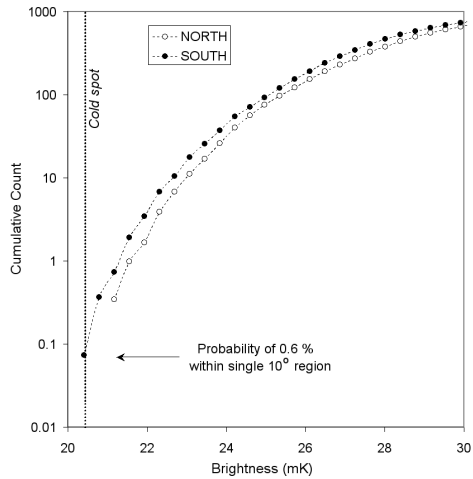}
\end{center}
\caption{The cumulative distribution, normalized to 1000, of median brightness levels (mK) in 3.4$^\circ$ sliding boxes of the NVSS images in two strips above $\vert$b$_{II} \vert>~$30$^\circ$ (see text). The minimum brightness (which is from the cold spot region) is indicated by a vertical line. }
\label{coldcum}
\end{figure}

\begin{figure}[t]
\begin{center}
\plottwo{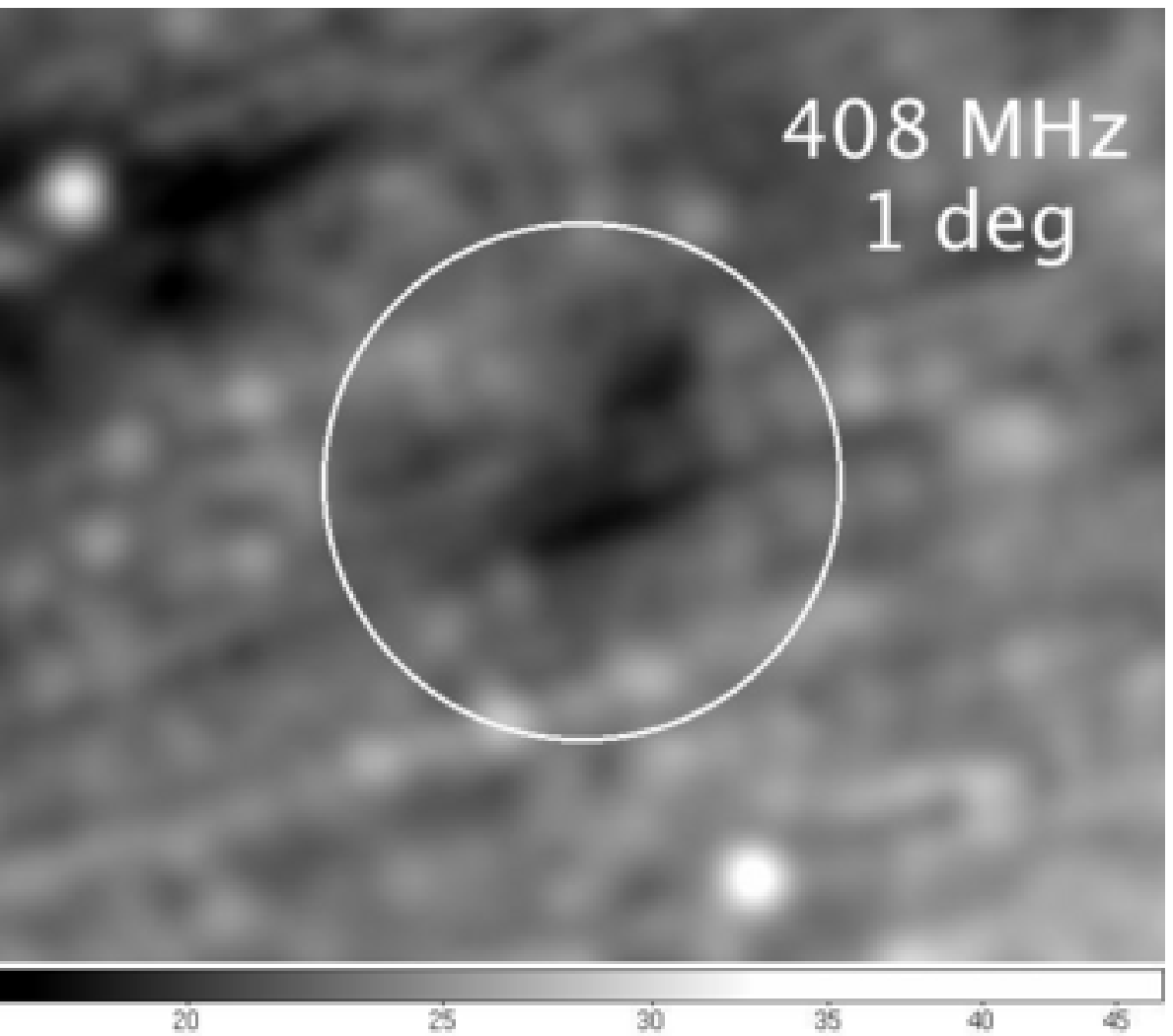}{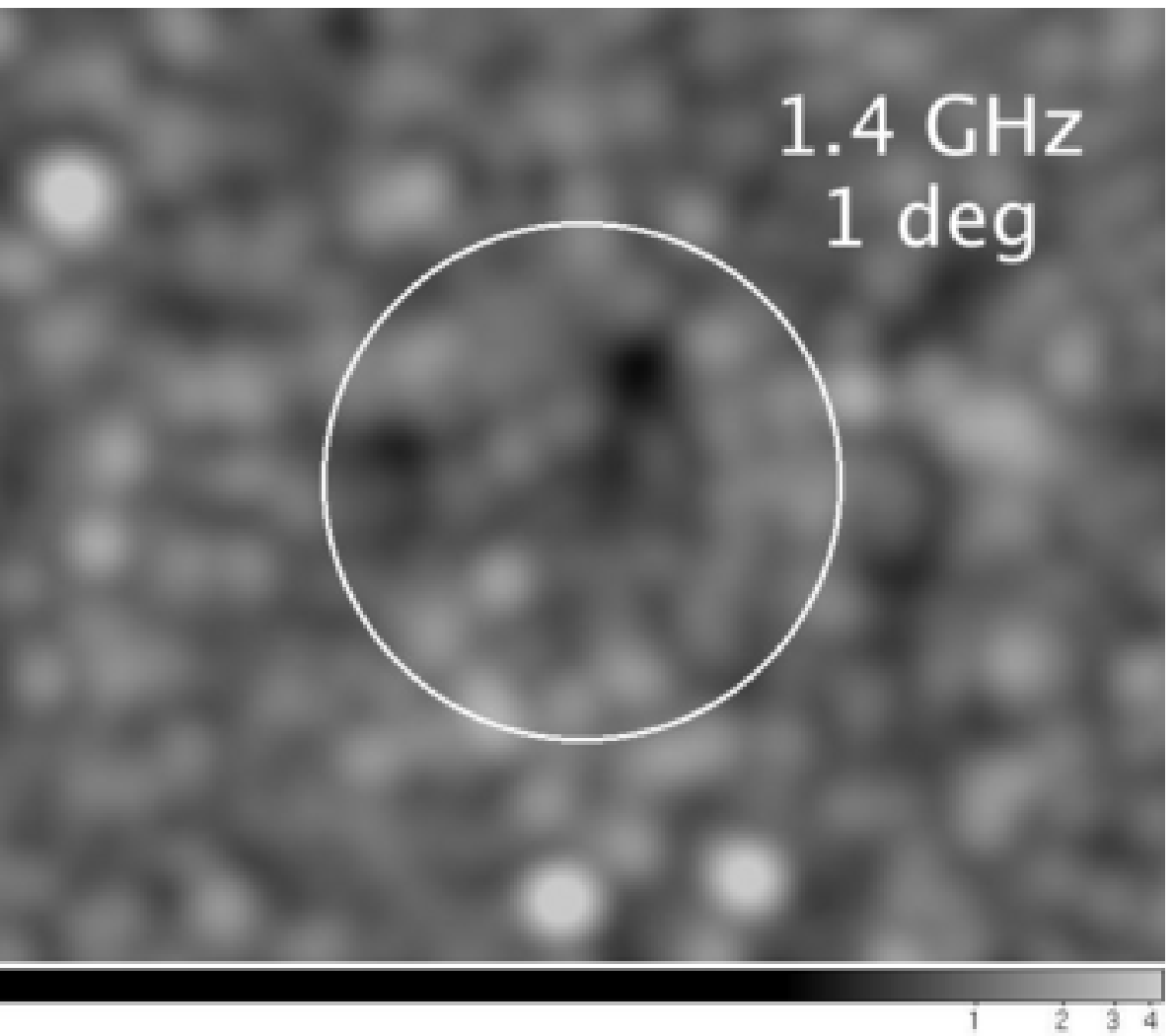}

\end{center}
\caption{18$^\circ$ fields, with 1$^\circ$ resolution, centered at  l$_{II}$, b$_{II}$ = 209$^\circ$, -57$^\circ$. Left:  408 MHz \citep{haslam81}. Right:  1.4 GHz \citep{nvss}.  A  10$^\circ$ diameter circle indicates the position and size of  the WMAP cold spot.}
\label{1d}
\end{figure}

\begin{figure}[t]
\begin{center}
\plotone{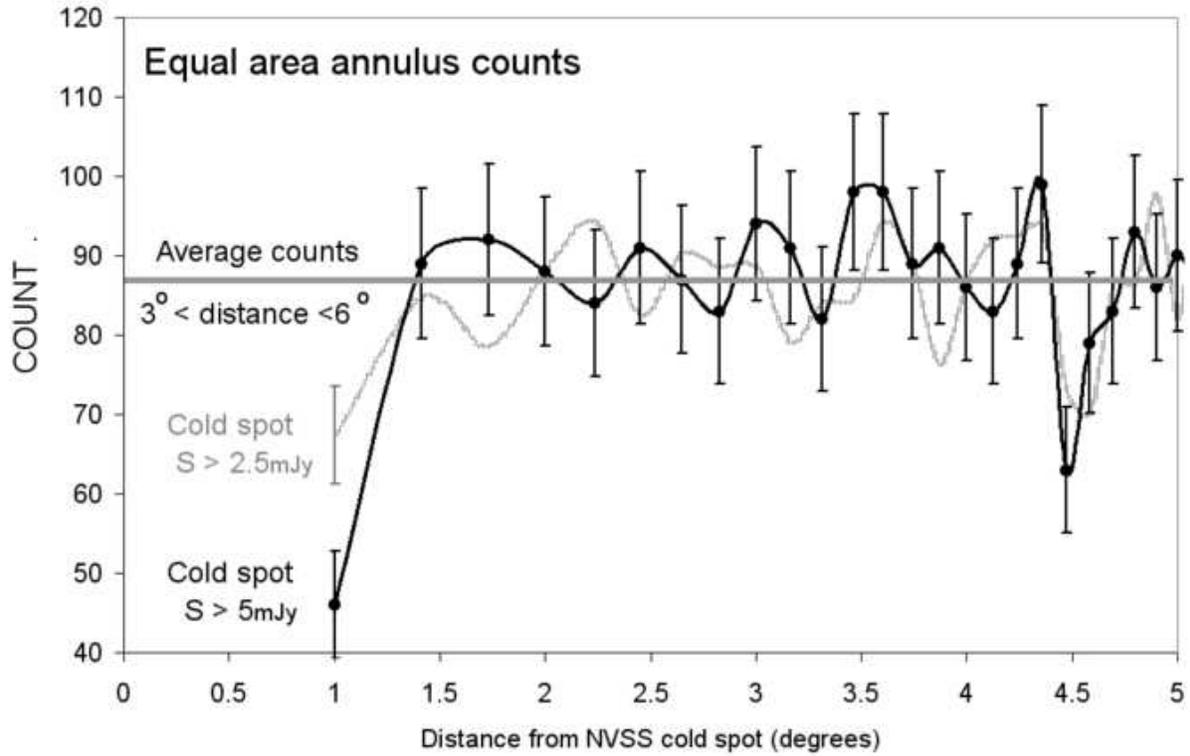}
\end{center}
\caption{Number of NVSS sources in 3.14 square degree annuli as a function of distance from the cold spot.  The counts axis refers to the results for counts of sources with S$>$5mJy; the grey line refers to counts for S$>$2.5mJy with those counts multiplied by 0.56. Each bin is independent. }
\label{density}
\end{figure}


\begin{thebibliography}

\bibitem[Bennett et al. (2003)]{1year03}
Bennett, C.L., et al. 2003, ApJS 148, 1

\bibitem[Cabre et al. (2006)]{cabre06}
Cabre, A., Gaztanaga, E., Manera, M., Fosalba, P. Castander, F. 2006, MNRAS 372, 23

\bibitem[Cayon, Jun \& Treaster (2005)]{cjt1}
Cayon, L., Jin J., Treaster, A. 2005, MNRAS 362, 826

\bibitem[Chiang \& Naselsky (2004)]{cn04} 
Chiang, L.-Y., Naselsky, P.D. 2006, IJMPD 15,1283C

\bibitem[Coles (2005)]{coles05} 
Coles, P. 2005, Nature 433, 248

\bibitem[Condon et al. (1998)]{nvss}
Condon, J. J., Cotton, W. D., Greisen, E. W., Yin, Q. F., Perley, R. A., Taylor, G. B., Broderick, J. J. 1998, AJ 115, 1693

\bibitem[Crittenden \& Turok (1996)]{crit1}
Crittenden, R. G., Turok, N. 1996, PRL 76, 575

\bibitem[Cruz et al. (2005)]{c05}
Cruz, M., Martinez-Gonzalez, E., Vielva, P., Cayon, L. 2005, MNRAS 356, 29

\bibitem[Cruz et al. (2006)]{c06}
Cruz, M., Tucci, M., Martinez-Gonzalez, E., Vielva, P. 2006, MNRAS 369, 57

\bibitem[Cruz et al. (2007)]{cruz07}
Cruz, M., Cayon, L., Martinez-Gonzalez, E., Vielva, P., Jin, J. 2007, ApJ 655, 11

\bibitem[de Oliveira-Costa et al. (2004)]{oliv}
de Oliveira-Costa A., et al. 2004, PhRevD 69, 3516

\bibitem[Giannantonio et al. (2006)]{gian06}
Giannantonio, T. et al. 2006, PhRvD 74, 352

\bibitem[Gumrukcuoglu, Contaldi \& Peloso (2006)]{pel06}
Gumrukcuoglu, A. E., Contaldi, C. R. \& Peloso, M. 2006, astro-ph/0608405

\bibitem[Haslam et al. (1981)]{haslam81} 
Haslam, C. G. T., Klein, U., Salter, C. J., Stoffel, H., Wilson, W.E., Cleary, M.N., Cooke, D.J., \& Thomasson, P. 1981, A\&A, 100,209


\bibitem[Hinshaw et al. (2006)]{3year06}
 Hinshaw, G., et al. 2006, ApJ, submitted
(astro-ph/0603451) 


\bibitem[Holdman, Mersini-Houghton \& Takahashi (2006a)]{hol6a}
 Holman, R.; Mersini-Houghton, L.; Takahashi, Tomo, 2006,
hep-th/0611223

\bibitem[Holdman, Mersini-Houghton \& Takahashi (2006b)]{hol6b}
 Holman, R.; Mersini-Houghton, L.; Takahashi, Tomo, 2006,
hep-th/0612142

\bibitem[Hoyle \& Vogeley (2004)]{hoyle}
Hoyle, F., Vogeley, M. S. 2004, ApJ 607, 751


\bibitem[Inoue \& Silk (2006a)]{in6a}
Inoue, K. T., Silk, J. 2006, ApJ, 648, 23 

\bibitem[Inoue \& Silk (2006b)]{in6b}
Inoue, K. T., Silk, J. 2006, astro-ph/0612347 

\bibitem[Jaffe et al. (2005)]{jaffe}
Jaffe, T. R., Banday A. J., Eriksen, H. K., Forski, K. M, Hansen, F. K. 2005, ApJ 629, L1


\bibitem[Jonas, Baart \& Nicolson (1998)]{rhodes98} 
Jonas, J., Baart, E. E., Necolson, G. D. 1998, MNRAS, 297, 997

\bibitem[Land \& Magueijo (2005)]{land}
Land, K., Magueijo, J.,2005,  PRL 95,071301

\bibitem[Liu \& Zhang (2005)]{lz05} 
Liu, X., \& Zhang, S. N. 2005, ApJ, 633,542

\bibitem[McEwen et al. (2007)]{MCE}
McEwen, J. D., Vielva, P., Hobson, M. P., Martinez-Gonzalez, E., \& Lasenby, A. N. 2007, MNRAS, in press


\bibitem[Pietrobon, Balbi \& Marinucci (2006)]{piet06}
Pietrobon, D., Balbi, A., Marinucci, D. 2006, PhysRevD 74, 352

\bibitem[Reich and Reich (1986)]{bonn}
Reich, W.  and Reich, P. 1986, A\&A Suppl.  63, 205

\bibitem[Tegmark et al. (2003)]{teg03} 
Tegmark, M., de Oliveira-Costa, A., Hamilton
A. 2003, Phys.Rev. D68, 123523.

\bibitem[Tojeiro et al. (2006)]{toj06} 
Tojeiro, R., Castro, P.G., Heavens, A.G.,
Gupta, S. 2006, MNRAS,365,265 

\bibitem[Vielva et al. (2004)]{v04}
Vielva, P., Martinez-Gonzalez, E., Varreiro, R. B., Sanz, J. L., Cayon, L. 2004, ApJ 609, 22

\bibitem[York et al. (2000)]{york}
York, D. G. et al. 2000, AJ 120, 1579

\bibitem[Zaldarriaga \& Seljak (1997)]{zald}
Zaldarriaga, M. \& Seljak, U. 1997, PhRvD 55, 1830

\end{thebibliography}
\end{document}